\documentclass[prl,twocolumn, aps,floats,showpacs,nofootinbib]{revtex4}
\usepackage{graphicx}
\usepackage{epsfig}
\usepackage{amsmath}
\usepackage{verbatim}
\usepackage{color}

\def\be{\begin{equation}}
\def\ee{\end{equation}}
\def\bea{\begin{eqnarray}}
\def\eea{\end{eqnarray}}
{
{

\def\gev{{\rm \,Ge\kern-0.125em V}}
\def\tev{{\rm \,Te\kern-0.125em V}}
\def\mev{{\rm \,Me\kern-0.125em V}}

\def\ren{\rm{ren}}

\def\half{\frac12}
\def\vecx{\vec{x}}
\def\vecy{\vec{y}}
\def\veck{\vec{k}}

\def\hatnp{\hat{n}^\prime}
\def\hatn{\hat{n}}

\begin{document}
\title{\bf
Power spectrum generated during inflation
}
\author{Mar Bastero-Gil}
\email{mbg@ugr.es}
\affiliation{Departamento de Fisica Teorica y del Cosmos, Universidad de Granada, 
Granada-18071, Spain}
\author{Arjun Berera}
\email{ab@ph.ed.ac.uk}
\affiliation{SUPA, School of Physics and Astronomy, University of Edinburgh, 
Edinburgh, EH9 3JZ, United Kingdom}
\author{Namit Mahajan}
\email{nmahajan@prl.res.in}
\author{Raghavan Rangarajan}
\email{raghavan@prl.res.in}
\affiliation{Theoretical Physics Division, Physical Research Laboratory,
Navrangpura, Ahmedabad 380 009, India}
\begin{abstract}
\noindent

Recently there have been differing viewpoints on how to evaluate the
curvature power spectrum generated during inflation.  Since the 
primordial curvature 
power spectrum is the seed for structure formation and provides a link between
observations and inflationary parameters, it is important to resolve any 
disagreements
over the expression for the power spectrum.  In this
article we discuss these differing viewpoints and indicate issues
that are relevant to both approaches.  
We then argue why the standard expression is 
valid.

\vskip 1cm
\noindent
\end{abstract}
\maketitle
\section{Introduction}
\noindent

In the standard inflationary Universe quantum fluctuations 
of the inflaton field give rise to a curvature perturbation that is
constant for modes outside the horizon.  This curvature perturbation is then
the seed for structure formation in the Universe.  For the inflaton field 
$\varphi$ given by
\begin{equation}
\varphi(\vec{x},t)=\frac{1}{(2\pi)^{3/2}}\int d^3 k \,
[ a_k \, \varphi_k(t) e^{i \veck.\vecx}
+ a_k^\dagger \,\varphi_k^*(t)  e^{-i \veck.\vecx}]
\end{equation}
the curvature perturbation
generated during inflation on superhorizon scales is given by
\begin{eqnarray}
\zeta_k &=&\frac{1}{3}\frac{\delta\rho(k)}{\rho+p}\\
&=& \frac{1}{3}\frac{V'(\varphi_0)\,\delta\phi(k)}{\dot\varphi_0^2}
\end{eqnarray}
where 
$\delta\varphi(k)^2= [k^3/(2\pi^2)]\,|\varphi_k|^2$ 
and $\varphi_0$ represents the classical homogeneous background.
For a very flat inflaton 
potential the inflaton can be taken to be massless and $\delta\varphi(k)=H/(2\pi)$
where $H$ is the Hubble parameter during inflation.  The curvature
power spectrum is defined as
$|\zeta_k|^2$. 

In a series of 
papers \cite{parker2007,parker2008,parker2009a,parker2009b,
parker2010a,parker2010b,parker2010c,parker2011}, it has been argued that the 
regularisation and renormalisation scheme adopted to make 
$\langle \varphi^2(x)\rangle=1/(2\pi)^3\int d^3k\,|\varphi_k|^2$ finite 
should also be applied when considering $\delta\varphi(k)^2$.  
Then in adiabatic regularisation the subtraction scheme applied to 
the
integrand in $\langle\varphi^2\rangle$ should be retained while obtaining
$\delta\varphi(k)^2$, and so
$\delta\varphi(k)^2=k^3/(2\pi^2)\,[|\varphi_k|^2 - |\varphi_K|^2]$, where
$\varphi_K$ is the adiabatic solution to the second order.
This then
modifies the power spectrum since
the subtraction scheme which cancels the contribution of high momentum modes 
in $\langle \varphi^2(x)\rangle$ 
also modifies the contribution of the superhorizon low momentum modes.  As argued
in Ref. \cite{parker2008,parker2009a,parker2009b}, this reduces the amplitude of 
the 
power spectrum for a massive inflaton, retains the scale free nature of the 
spectrum, modifies the tensor-scalar ratio $r$,
and allows for the compatibility of
quartic chaotic inflation with data.  

However, it was argued in Ref. \cite{durrer2009} that while 
the fluctuation mode functions are
constant outside the horizon the adiabatic solution 
is not and so 
the power
spectrum then depends on the time after horizon crossing
at which the power spectrum is evaluated.
It was also argued that different adiabatic subtraction schemes gave 
different results.  It was therefore concluded that one should carry
out adiabatic subtraction only for high momentum modes.

The above result was countered in Ref. \cite{parker2010a} by arguing that 
adiabatic regularisation required subtracting the adiabatic solution for all
modes, not just high momentum modes.  The authors further argued that their adiabatic
subtraction scheme differed from that in Ref. \cite{durrer2009}, and that their
scheme agreed with de Witt-Schwinger renormalisation (in momentum space in the
massless limit) which identifies counterterms without invoking any adiabatic
condition.

Ref. \cite{durrer2011} then argued that the de Witt-Schwinger expansion 
is relevant for large momentum modes but
is not valid for superhorizon modes that leave the horizon.  This was further
countered by Ref. \cite{parker2011} wherein it was re-emphasised that adiabatic
subtraction must be applied to all modes,
that the energy momentum tensor and $\langle\varphi^2\rangle$
require mode subtractions at the 4th and 2nd order respectively, and that
the adiabatic solution at the appropriate order
need not approximate the solution
for all momenta.

Since the curvature power spectrum is an essential ingredient in the process of
extracting early Universe parameters from current observations, it is important 
that the
above issues be resolved and that there is clarity on what is the appropriate
expression for the power spectrum.
Below we comment on some issues related to both viewpoints on obtaining the
power spectrum and then present arguments as to why the standard expression in the
literature is appropriate.  

\section{The power spectrum}
The argument in Refs. \cite{durrer2009,durrer2011} on applying the subtraction 
scheme only to high momentum modes is equivalent to introducing 
a time dependent cutoff such as  
$\Theta(k-aH)$ to subtract only high momentum
modes while calculating $\langle\varphi^2(x)\rangle$.  
(Refs. \cite{durrer2009,durrer2011}
actually calculate $\langle Q^2(x)\rangle$, where $Q$ is the Mukhanov-Sasaki
variable.)  Now for a rigid spacetime ignoring metric perturbations
the equation of motion for $\varphi_k$ implies 
\begin{equation}
\dot\rho_k = - 3H (\rho_k + p_k)\,.
\end{equation}
Integrating over all $k$ modes then gives
\begin{equation}
\dot\rho_\varphi = -3 H (\rho_\varphi + p_\varphi) \,.
\label{energyeq}
\end{equation}
But if we replace $\rho_\varphi$ and $p_\varphi$ by renormalised quantities 
$\rho_{\ren}$ and $p_{\ren}$ with the contribution of 
high momentum 
modes cut off at $k = a(t) H$, then this time dependent
cutoff spoils the  equality above because the time derivative on the left
hand side of Eq. (\ref{energyeq}) acts on the cutoff too.  
\begin{equation}
\dot\rho_{\ren} = 
\frac{d}{dt} \int\frac{d^3k}{(2\pi)^{3/2}} [\rho_k - \Theta(k-aH) \rho_K]
e^{i\veck.\vecx}
\end{equation}
and
\begin{align}
-3H (\rho_{\ren} + p_{\ren})
= - 3 H 
\int &\frac{d^3k}{(2\pi)^{3/2}}  
[\rho_k - \Theta(k-aH) \rho_K
\nonumber\\
  &+ p_k - \Theta(k-aH) p_K]e^{i\veck.\vecx}
\end{align}
where the subscript $K$ refers to the adiabatic solution.  With the adiabatic 
solution
cancelling (to the relevant adiabatic order) the high momentum contribution we 
then get
\begin{align}
\dot\rho_{\ren} 
&= \frac{d}{dt} \int_0^{a(t)H}  \frac{d^3k}{(2\pi)^{3/2}} \,\rho_k 
e^{i\veck.\vecx}
\nonumber\\
&\ne
-3H (\rho_{\ren} + p_{\ren})
\nonumber\\
&= - 3 H 
\int_0^{a(t)H} \frac{d^3k}{(2\pi)^{3/2}} \,[\rho_k   + p_k ]e^{i\veck.\vecx}
\end{align}
because of the contribution of the time derivative acting on the upper limit of 
the first
integral.  This suggests that a regularisation prescription, as proposed by
Refs. \cite{durrer2009,durrer2011}, that only subtracts
the high momentum modes is not appropriate.

But
one may now
question whether regularisation and renormalisation itself are relevant for the 
power spectrum, as insisted on by Refs. 
\cite{parker2007,parker2008,parker2009a,parker2009b,
parker2010a,parker2010b,parker2010c,parker2011}.  After all,
the curvature power spectrum depends on 
$\delta\varphi(k)$ and not $\langle\varphi^2(x)\rangle$,
and it is the latter that involves
the divergent integral over $k$.  
This issue can be resolved by
identifying
the quantity that enters
in physical observables or in expressions derived from physical observables.
Let us consider the cosmic microwaved background (CMB) 
temperature anisotropy variable
\begin{equation}
C_l=\frac{1}{4\pi} \int d^2 \hat{n} \,d^2\hat{n}^\prime\, 
P_l(\hatn.\hatnp)\,\langle \Delta T(\hatn)\Delta T(\hatnp)\rangle
\end{equation}
where $\Delta T(\hatn) = T(\hatn)-T_0$ represents the difference in temperature 
of the CMB in a direction $\hatn$ from the mean temperature $T_0$.
$\langle \Delta T(\hatn)\Delta T(\hatnp)\rangle$ above is obtained from 
observations.
\footnote{More precisely, we measure $\Delta T(\hatn)\Delta T(\hatnp)$.  The
difference gives
rise to cosmic variance \cite{wbergcosmology}, which we ignore here.
}
Then using 
\begin{equation}
\left(\frac{\Delta T(\hatn)}{T_0}\right)_{SW}= \frac{1}{3}\,\delta\phi(\hatn r_L)
\,,
\end{equation}
where $r_L$ is the distance to the surface of last scattering, 
$\delta\phi$ is the perturbation in the gravitational potential and $SW$ refers to
the Sachs-Wolfe effect, we get
\begin{align}
C_l\, &\sim \hspace{0.3cm}...\hspace{1cm}\langle \delta\phi(\hatn r_L)\,\delta\phi
(\hatnp r_L)\rangle
\nonumber\\
&\sim \hspace{0.3cm}...\hspace {1cm}
\int d^3 q\,d^3 q' e^{i \vec{q}.\hatn r_L}\,e^{i \vec{q}'.\hatnp r_L}
\langle \delta\phi_{\vec{q}}\,\,\delta\phi_{\vec{q}'}\rangle
\nonumber\\
&\sim \hspace{0.3cm}...\hspace {1cm}
\int d^3 q\,d^3 q' e^{i \vec{q}.\hatn r_L}\,e^{i \vec{q}'.\hatnp r_L}
P_\phi(q) \,\delta(\vec{q}+\vec{q}')
\nonumber\\
&\sim \hspace{0.3cm}...\hspace {1cm}
\int d^3 q e^{i \vec{q}.\hatn r_L}\,e^{-i \vec{q}.\hatnp r_L}
P_\phi(q) \,,
\label{Cl}
\end{align}
where $\langle \delta\phi_{\vec{q}}\,\delta\phi_{\vec{q}'}\rangle
=P_\phi(q) \,\delta(\vec{q}+\vec{q}')$ and $P_\phi(q)$ is the power spectrum
associated with $\delta\phi$.  Thus we see that 
it is the coordinate space correlation function of the gravitational potential 
perturbation that is primary.  Since the gravitational
potential perturbation is related to quantum fluctuations of the inflaton
we would argue that the relevant quantity for physical observables is the
inflaton correlation function in coordinate space,
and this must be renormalised and finite.  Then, as argued in
Refs. \cite{parker2007,parker2008,parker2009a,parker2009b,
parker2010a,parker2010b,parker2010c,parker2011},
the power 
spectrum should reflect the renormalisation prescription for 
the coordinate space correlation function for the inflaton field.

For a massless 
scalar field 
\begin{equation}
\varphi_k(t)=\frac{iH}{(2k^3)^\half}[1 + i k \tau]\exp(-ik\tau)
\end{equation}
where $\tau=-1/[a(t)H]$.  Then
\begin{equation}
\langle\varphi^2\rangle=\frac{1}{(2\pi)^3} \int d^3k 
\left[\frac{1}{2ka^2} + \frac{H^2}{2k^3} \right]
\label{phisq}
\end{equation}
Refs. \cite{parker2007,parker2008,parker2009a,parker2009b,
parker2010a,parker2010b,parker2010c,parker2011} define the power spectrum 
using the adiabatically
regularised $|\varphi_k|^2$ needed for regularising $\langle\varphi^2(x)\rangle$.
Such a prescription would eliminate {\it {both}} the terms in the integrand
of Eq. (\ref{phisq}).  For a scalar field of mass $m$ the final power spectrum  
would be driven by the scale $m$ rather than $H$ \cite{parker2008}.

But Eq. (\ref{Cl}) indicates that $C_l$ actually depends on the correlation 
function of the inflaton
at two different points in space.  
So we would argue that $\langle\varphi(\vecx,t)\varphi(\vecy,t)\rangle$ 
is the relevant 
quantity
to be used to obtain the power spectrum for $\varphi$, and so the power 
spectrum should reflect the renormalisation prescription, if any, for 
$\langle\varphi(\vecx,t)\varphi(\vecy,t)\rangle$,  rather than 
for $\langle\varphi^2(x)\rangle$.
Now
\begin{equation}
\langle\varphi(\vecx,t)\varphi(\vecy,t)\rangle=\frac{1}{2\pi^2} \int  dk \,k^2
\left[\frac{1}{2ka^2} + \frac{H^2}{2k^3}\right] 
\frac{\sin[k|\vecx-\vecy|]}{k|\vecx-\vecy|}
\label{phixphiy}
\end{equation}
This quantity does not require renormalisation as the sine function makes the
integral ultraviolet finite.  
\footnote{
Note that the first term is present even in flat spacetime, and is finite and 
equal to the equal time Feynman propogator for a massless scalar field, namely 
$i/(4\pi^2|\vecx-\vecy|^2)$, 
in Minkowski spacetime \cite{greinerreinhardt}.  
}
Therefore there will be no need of adiabatic subtraction and hence no 
modification of the
integrand.  Then associating $\delta\varphi(k)$ with
the expression in brackets in Eq. (\ref{phixphiy})
we get the standard expression for
$\delta\varphi(k)=H/(2\pi)$,  
and thus for the primordial curvature power spectrum.
Note that
if we define the power spectrum using 
Eq. (\ref{phixphiy}) then both the terms in the brackets 
will be included but only the second term 
contributes in the large wavelength limit, $k\ll aH$.
  
We believe that the above prescription might be the
appropriate way of obtaining the power spectrum generated during inflation. 
The power spectrum is also 
not time dependent as in the prescription of Refs. 
\cite{parker2007,parker2008,parker2009a,parker2009b,
parker2010a,parker2010b,parker2010c,parker2011}.  

In the literature different authors define the power spectrum 
using either the integrand of $\langle\varphi^2(x)\rangle$ or 
of $\langle\varphi(\vecx,t)
\varphi(\vecy,t)\rangle$. If one is ignoring renormalisation of these 
quantities then both approaches give the same momentum space power spectrum. 
But, as we clarify above,
the spatial corrrelation function enters in the expression for
physical observables like $C_l$ and so 
one must consider renormalised quantities in coordinate space, and hence in
momentum space too, as argued in 
Refs. \cite{parker2007,parker2008,parker2009a,parker2009b,
parker2010a,parker2010b,parker2010c,parker2011}.
However, the spatial correlation function that is relevant is 
$\langle\varphi(\vecx,t)\varphi(\vecy,t)\rangle$, not 
the divergent $\langle\varphi^2(x)\rangle$ which is considered in 
Refs. \cite{parker2007,parker2008,parker2009a,parker2009b,
parker2010a,parker2010b,parker2010c,parker2011},
and the former quantity 
does not require regularisation.  We thus
get the standard expression for the power spectrum.

We must add here that we have only temporarily set aside the necessity of 
renormalisation of $\langle\varphi^2(x)\rangle$.  In an interacting theory, 
our prescription above is relevant
for calculating the power spectrum only to lowest order.  
For example, in a $\lambda\varphi^4$ theory
\begin{equation}
\langle\varphi(x)\varphi(y)\rangle_{\rm int}
= \langle\varphi(x)\varphi(y)\rangle + 
i \lambda\int d^4z\,\langle \varphi(x)\varphi(y)\varphi^4(z)\rangle
\end{equation}
and the second term will be proportional to $\langle\varphi^2(z)\rangle$, which 
will require renormalisation. 
(Note, however, that a cubic self interaction will not require such 
renormalisation
of the correlation function.)
Renormalisation of $\langle\varphi^2(x)\rangle$ will also be needed for
obtaining the renormalised energy momentum tensor in free and interacting
field theories.  
\footnote{
We point out a subtlety here.  We have been using
$\langle\varphi^2(x)\rangle$ for $\langle\varphi(x)\varphi(x)\rangle$
while discussing the power spectrum.
But this is a slight abuse of notation.  Technically speaking,
$\varphi^2(x)$ is a composite operator and
\begin{equation}
\langle\varphi^2(x)\rangle=\langle\varphi(x)\varphi(x)\rangle +\,...\,,
\end{equation}
where we have used
the operator product expansion. 
In fact, for the energy momentum tensor it is the quantity on the lhs that 
is needed.
}

We mention in passing that the expression for
$\langle\varphi(\vecx,t)\varphi(\vecy,t)\rangle$
has an infrared divergence just like $\langle\varphi^2(x)\rangle$.
However for realistic inflation models the inflaton may have a mass, albeit
small, or the mass 
may even be generated non-perturbatively \cite{Beneke:2012kn},
or inflation may be preceded by a radiation dominated era, 
which should remove the infrared divergence.

\section{Conclusion}

In conclusion, in this article 
we have discussed two differing viewpoints on obtaining the
power spectrum generated during inflation.  We point out that subtracting only 
the
contribution of high momentum modes to $\langle\varphi^2\rangle$, 
as suggested by
Refs. \cite{durrer2009,durrer2011}, 
may not be appropriate as one does not obtain the standard
energy equation for renormalised quantities.  However we also point out that, 
keeping in mind physical observables, it is more relevant  
to obtain the power spectrum
from 
$\langle\varphi(\vecx,t)\varphi(\vecy,t)\rangle$, 
rather than 
from $\langle\varphi^2(x)\rangle_{\ren}$ as 
suggested in Refs. 
\cite{parker2007,parker2008,parker2009a,parker2009b,
parker2010a,parker2010b,parker2010c,parker2011}.
Our prescription then gives 
the standard expression
for the power spectrum and thereby validates the results in the literature based
on this expression.

After completing this work, we came across Ref. \cite{finellietal} which
contains arguments similar to ours pertaining to the choice of the
correlation function, and divergences at higher order.  
It is surprising that their
arguments have not been emphasised in the literature.

{\bf Acknowledgements:} AB would like to thank Physical Reserach Laboratory, 
Ahmedabad, India for support
and hospitality during a visit when this work was initiated, 
and STFC, UK for support.
M.B-G. is partially supported by MICINN (FIS2010-17395) and
``Junta de Andaluc\'ia" (FQM101).


\begin{thebibliography}{99}

\bibitem{parker2007} 
  L.~Parker,
  hep-th/0702216 [HEP-TH].


\bibitem{parker2008} 
  I.~Agullo, J.~Navarro-Salas, G.~J.~Olmo and L.~Parker,
  Phys.\ Rev.\ Lett.\  {\bf 101}, 171301 (2008)
  [arXiv:0806.0034 [gr-qc]].

 \bibitem{parker2009a}
  I.~Agullo, J.~Navarro-Salas, G.~J.~Olmo and L.~Parker,
  Phys.\ Rev.\ Lett.\  {\bf 103} (2009) 061301
  [arXiv:0901.0439 [astro-ph.CO]].
  
 \bibitem{parker2009b} 
  I.~Agullo, J.~Navarro-Salas, G.~J.~Olmo and L.~Parker,
  Gen.\ Rel.\ Grav.\  {\bf 41}, 2301 (2009)
  [Int.\ J.\ Mod.\ Phys.\ D {\bf 18}, 2329 (2009)]
  [arXiv:0909.0026 [gr-qc]].
  
 \bibitem{parker2010a} 
  I.~Agullo, J.~Navarro-Salas, G.~J.~Olmo and L.~Parker,
  Phys.\ Rev.\ D {\bf 81}, 043514 (2010)
  [arXiv:0911.0961 [hep-th]].
  
\bibitem{parker2010b} 
  I.~Agullo, J.~Navarro-Salas, G.~J.~Olmo and L.~Parker,
  J.\ Phys.\ Conf.\ Ser.\  {\bf 229}, 012058 (2010)
  [arXiv:1002.3914 [gr-qc]].
  
  
 \bibitem{parker2010c} 
  I.~Agullo, J.~Navarro-Salas, G.~J.~Olmo and L.~Parker,
  arXiv:1005.2727 [astro-ph.CO].
  
 \bibitem{parker2011} 
  I.~Agullo, J.~Navarro-Salas, G.~J.~Olmo and L.~Parker,
  Phys.\ Rev.\ D {\bf 84}, 107304 (2011)
  [arXiv:1108.0949 [gr-qc]].
  
\bibitem{durrer2009} 
  R.~Durrer, G.~Marozzi and M.~Rinaldi,
  Phys.\ Rev.\ D {\bf 80}, 065024 (2009)
  [arXiv:0906.4772 [astro-ph.CO]].
  
\bibitem{durrer2011} 
  G.~Marozzi, M.~Rinaldi and R.~Durrer,
  Phys.\ Rev.\ D {\bf 83}, 105017 (2011)
  [arXiv:1102.2206 [astro-ph.CO]].
  
  

\bibitem{wbergcosmology} 
  S.~Weinberg,
  ``Cosmology'',
  Oxford, UK: Oxford Univ. Pr. (2008) (see Sec. 2.6).
  
\bibitem{greinerreinhardt}
W. Greiner and J. Reinhardt, ``Quantum Electrodynamics",
2nd edition, Springer (1994) (see Ex. 2.5)




\bibitem{Beneke:2012kn}
  M.~Beneke and P.~Moch,
  Phys.\ Rev.\ D {\bf 87}, 064018 (2013)
  arXiv:1212.3058 [hep-th].

\bibitem{finellietal} 
  F.~Finelli, G.~Marozzi, G.~P.~Vacca and G.~Venturi,
  Phys.\ Rev.\ D {\bf 76}, 103528 (2007)
  [arXiv:0707.1416 [hep-th]].

  
\end{thebibliography}
\end{document}